\documentstyle[12pt,epsf]{article}
\def\gsim{\;\raise0.3ex\hbox{$>$\kern-0.75em\raise-1.1ex\hbox{$\sim$}}\;}
\def\lsim{\;\raise0.3ex\hbox{$<$\kern-0.75em\raise-1.1ex\hbox{$\sim$}}\;}
\newcommand{\be}{\begin{equation}}

\newcommand{\ee}{\end{equation}}
\newcommand{\bea}{\begin{eqnarray}}
\newcommand{\eea}{\end{eqnarray}}
\newcommand{\bt}{\begin{tabular}}
\newcommand{\et}{\end{tabular}}
\newcommand{\ba}{\begin{array}}
\newcommand{\ea}{\end{array}}

\begin{document}

\thispagestyle{empty}

\setcounter{page}{0}

{}\hfill{DSF$-$05/2006}


\vspace{1truecm}

\begin{center}
{\Large \bf Majorana and the investigation of infrared spectra of
ammonia.}
\end{center}

\bigskip\bigskip

\begin{center}
{\bf E. Di Grezia$^{1,2,a}$.}

\vspace{.5cm}

$^!$ {\it Dipartimento di Scienze Fisiche, Universit\`{a} di
Napoli ``Federico II'' \\ Complesso Universitario di Monte S.
Angelo, Via Cinthia, I-80126 Napoli, Italy}

$^2$ {\it Istituto Nazionale di Fisica Nucleare, Sezione di
Napoli, Complesso Universitario di Monte S. Angelo, Via Cinthia,
I-80126 Napoli, Italy}

$^a$ e-mail address: Elisabetta.Digrezia@na.infn.it

\end{center}

\bigskip\bigskip\bigskip

\vspace{3cm}
\begin{abstract}
\noindent An account is given on the first studies on the physics
of ammonia, focusing on the infrared spectra of that molecule.
Relevant contributions from several authors, in the years until
1932, are pointed out, discussing also an unknown study by
E.Majorana on this topic.
\end{abstract}

${}$ \\


\newpage

\section{Introduction}
Because of the intensity and richness of its spectrum, ammonia has
played a great role in the development of microwave spectroscopy.
It has provided a large number of observable lines on which to try
both experimental techniques and the theory. $NH_3$ provides the
simplest and most thoroughly worked out example of a class of
spectra which occupied and puzzle microwave spectroscopists for
many years. In the paper of $1932$ Fermi \cite{Fm} discusses the
influence of the ammonia molecule's rotation on the doubling of
its levels. This doubling originates -according Dennis and Hardy
\cite{dh}- in the oscillation by which the nitrogen atom crosses
the plane determined by the three hydrogens, i.e., due to
inversion respect the plane of the three atoms of $H$ influenced
by the rotation of molecule and he compared the theoretical
results with the experimental results and he found accords
 (inversion problem). This paper on $NH_3$, together with other
three articles on the accidental degeneracy of the carbon dioxide
molecule's frequencies of oscillation on the Raman effect in
crystals, constitutes a series of investigations from the period
1931-33 in which Fermi attempted to explain various molecular
phenomena. His interest in these studies is reflected in the book
"Molecules and crystals" \cite{Fm1}. Fermi's interest in this type
of problem was to give a quantitative explanation to experimental
observation at the center of Institutes by Rasetti, which in the
period 1929-1930 studied the Raman effect in diatomic gas $O_2$,
$N_2$. In particular the paper on the $NH_3$ molecule is connected
to experiments conducted in Rome during the same period by E.
Amaldi on the Raman Effects with theoretical contribute by G.
Palczek \cite{[FN16]}, and theoretical research by G. Placzek and
E. Teller \cite{[FN19]} on molecular spectra for $CO_2$ and
$NH_3$.
\\
In reality W.W. Coblentz \cite{[FN30]} in 1905 investigated the
positions and fine structure of the infrared bands of polyatomic
molecules $CO_2$, $NH_3$. Coblentz observed two very intense bands
at $\lambda = 10.7 \mu, 6.14 \mu$, with a considerable weaker band
at $\lambda=2.97 \mu$.
\\
Lately K. Schierklok \cite{[FN31]} has re-examined the ammonia
infra-red (IR) spectrum and in addition to the bands found by
Coblentz, he has found a band at $\lambda=2.22 \mu$. Beyond this
he found two bands at $\lambda=1.94 \mu, 1.49 \mu$ whose intensity
was about half that of the two previous bands; so Schierklok
observed six bands.

Historically the study of systems with several atoms, combining or
not to form a molecule, has interested chemists for many years
through the rules of valence. But only at the beginning of $"900$
the physicists interpret these rules in the light of quantum
mechanics and the behavior of the constituents of atom with the
spectral analysis of radiation emitted by the atom. Heitler and
London \cite{hl} connected the valence in the formation of
homopolar diatomic molecules with symmetry character of wave
functions of the outer electrons in each atom. The problem of the
vibration groups of atoms possessing geometric symmetry has been
considered for the first time by C. J. Brester \cite{br}. A number
of models representing particular molecules have been treated
making use of various assumptions to obtain the potential energy
function. Historically the first example is that of $CO_2$
\cite{Co}. Hund \cite{[FN32]} and Kornfeld \cite{Ko} examined the
spectra of $H_2O, H_2S, CO_3$ ion, $NH_3$ \cite{nh}. Dennison
\cite{[FN33]} found the normal vibrations for models of $NH_3$ and
$CH_4$ assuming the forces to be central and Nielsen made a like
treatment of the  $CO_3$ ion. In all these investigations, the
molecule was assumed to have a certain geometry symmetry in its
equilibrium configuration. So another more simple way to analyze
these models is to use the theory of vibrations \cite{vib}. In
particular our interest will be on the $NH_3$- molecule which
Coblentz in $1905$ will start to study. In the present article we
will investigate how the $NH_3$- molecule, in particular, has been
studied qualitatively by character of the vibration of symmetrical
polyatomic molecules through the  theory of vibrations
\cite{[FN32]}, \cite{[FN33]}, approximatively using the
Wentzel-Kramers-Brillouin method of approximation. And a
quantitative analysis with exact solution for a two-minima problem
of the ammonia molecule solving
secular equation.\\
We will analyze the historical development of important works in
studying molecular spectra by means of the quantum theory, and in
obtaining information about the structure of the molecule through
an examination of the positions and fine structure of the IR
bands. Then we will present briefly the experiments in
observations of vibrational and rotational transitions in the
cases of gas $NH_3$ and the comparison of theoretical and
experimental amounts of vibrational frequencies.

Different activities of the Fermi group at the Physics Institute
in Rome from 1930 to 1932 was devoted to this subject. In this
contest is the contribute of Majorana.

\section{Theoretical analysis of ammonia spectra.}
\subsection{A brief introduction to the Theory of Vibrations.}
The classical theory of vibrations about an equilibrium
configuration has developed from Galileo's study of small
oscillations of a pendulum. In the first half of the eighteenth
century Brook Taylor, D'Alambert, Euler, Daniel Bernoulli
investigated the vibrations of a stretched cord. In 1753 Bernoulli
enunciated the principle of the resolution of all compound types
of vibration into independent modes. In $1762-1765$ Lagrange gave
the general theory of the vibrations of a dynamical system with a
finite number of degrees of freedom. One considers a vibrating
system defined by its kinetic energy $T$ and its potential energy
$V$ and its position is specified by a set of coordinates $(q_1,
q_2,..,q_n)$, giving the displacements from equilibrium. The
problem of vibrations around an equilibrium configuration is to
solve Lagrangian equations of motion in which the kinetic $T$ (a
positive definite form with $|a_{ij}|\neq 0$) and a potential
energies $V$ (Taylor expansion in powers of $q_1, q_2, ...,q_n$)
are homogenous quadratic forms in velocities and coordinates
respectively, with constant coefficients:
\begin{equation}
T = \frac{1}{2} \, \left( a_{11}\dot{q}_1^2 \, + ...  \,
a_{22}\dot{q}_2^2 \,+ \cdots + a_{nn}\dot{q}_n^2 \, +  \, 2
a_{12}\dot{q}_1 \dot{q}_2 \, +  \, 2 a_{13}\dot{q}_1 \dot{q}_3 \,
+\,+  \, 2 a_{n-1}\dot{q}_{n-1} \dot{q}_n \, \right) \label{1}
\end{equation}
\begin{equation}
V = \frac{1}{2} \, \left( b_{11}q_1^2 \, + b_{22}q_2^2 +...  \, +
b_{nn}q_n^2 \, +  \, 2 b_{12}q_1 q_2 \, + \,2 b_{13}q_1 q_3 \, + 2
b_{n-1}q_{n-1} q_n \right) \label{2}
\end{equation}
The equation of motion are:
\begin{equation}
\frac{d}{dt}\left(\frac{\partial T}{\partial \dot{q}_r}\right) =
-\frac{\partial V}{\partial q_r}\,\,\,\,\,\,\,(r=1,2,\cdots n)
\end{equation}
If $T$ and $V$ has the form (\ref{1}), (\ref{2}) (following the
method of Jordan \cite{Jr}), it is always possible to find a

linear transformation of coordinates $q_i=\sum_{k=1}^{n}c_{ik}x_k$
such that the kinetic and potential energies, expressed in terms
of the new coordinates, called {\em normal (principal)
coordinates}, have the form:
\begin{equation}
T = \frac{1}{2} \, \left( \dot{x}_1^2 \, + ...  \, \dot{x}_2^2 \,
+ \cdots + \dot{x}_n^2 \, \right) \label{3}
\end{equation}
\begin{equation}
V = \frac{1}{2} \, \left( \lambda_{1}x_1^2 \, + \lambda_{2}x_2^2
+... \, + \lambda_{n}x_n^2  \right) \label{4}
\end{equation}
where the constants $\lambda_1,\cdots,\lambda_n$, which occur as
coefficients of the squares of $x_k$ in $V$, are the $n$ distinct
or multiple roots of the determinant $det (a_{ik}\lambda - b_{ik})
=0$, and $a_{ik}$, $b_{ik}$ are the coefficients in the original
expressions of $T$ and $V$ energies. The Lagrangian equation of
motion is therefore:
\begin{equation}
\ddot{x}_r + \lambda_rx_r =0 \,\,\,\,\, r=1,2,\cdots,n.
\end{equation}
Thus the classical theory of small oscillations shows that system
will vibrate as an aggregation of $n$ independent mode of
vibration of the system, provided the corresponding constant
$\lambda_r$ is positive (stable equilibrium configuration), with
normal or characteristic frequencies $\nu_i =
\lambda_i^{1/2}/2\pi$. Moreover every conceivable vibration of the
system may be regarded as the superposition of $n$ independent
normal vibrations according Daniel Bernoulli's principle
\cite{bern}

\subsection{Ammonia molecule analysis with Theory of Vibrations.}
The vibration spectra of polyatomic molecules, in particular of
$NH_3$, has been investigated in great details both theoretically
and experimentally. For this system one has a number $s=4$ of
atomic nuclei which one assumes to have a possible equilibrium
position. Dealing with the internal or vibrational degrees of
freedom, the whole system has $n= 3s-6 =6$ degrees of freedom. The
ammonia molecule is like a one-dimensional system of a particle
moving in a potential field consisting of two equal minima and was
first treated qualitatively by Dennison and Hund \cite{[FN32]},
\cite{[FN33]}. Dennison and Hund assumed that the behavior of the
nuclei in the neighborhood of their equilibrium positions may be
described by means of central forces acting between them in the
case of a polyatomic molecule with certain limitations in regard
to the character of the equilibrium of the system.
\\
The assumption for a molecule of the type $XY_3$ is that in the
normal state of the molecule the $X-$atom is equidistant from each
of the $Y-$atoms which themselves lie at the corners of an
equilateral triangle. It is further assumed that the $X-$atom does
only four frequencies, as indeed will any model which posses an
axis of symmetry (Hund) and so $X$ is at the apex of a regular
pyramid with an equilateral triangle as a base. Experimentally
four fundamental $\nu$ are found (without axial symmetry are found
six fundamental frequencies).
\\
Then they assumed that the four independent active frequencies are
four fundamental absorption bands because of their fine structure.

Their study of $NH_3$ was motivated from IR spectroscopy  measures
and Raman spectra for polyatomic molecules $CO_2, N_2O, NH_3,
CH_4, C_2H_4$ during the period $1905-1935$. Dennison and Hund,
separately, showed, for molecules $H_2O, NH_3, CH_4$, that the
vibrational levels which lie below the potential maximum occur in
pairs. To find the normal vibrations they used the wave mechanical
treatment of vibration spectrum of $NH_3$-molecule and to obtain
their properties they investigated the geometric symmetry of
$NH_3$ in its equilibrium configuration. Let there be chosen a set
of coordinates $q_1,..,q_6$ giving the displacements from
equilibrium. In considering the system either in classical
mechanics or in wave mechanics, the first step is to find the
Hamiltonian. To the approximation in which the motions of the
atoms are small compared with the inter-atomic distances, the
system may absorb or emit radiation with a series of frequencies.
These frequencies are the so-called normal frequencies and may be
computed with the classical theory of small oscillations (theory
of vibrations \cite{vib} we have summarized in the above section),
for which in first approximation the kinetic and potential
energies assume a simple form:
\begin{equation}
T = \frac{1}{2} \, \left( a_{11}\dot{q}_1^2 \, + ...  \, +
a_{66}\dot{q}_6^2 \, +  \, 2 a_{12}\dot{q}_1 \dot{q}_2 \, +\,
....\right)
\end{equation}
\begin{equation}
V = \frac{1}{2} \, \left( b_{11}q_1^2 \, + ...  \, + b_{66}q_6^2
\, +  \, 2 b_{12}q_1 q_2 \, +\, .... \right)
\end{equation}
where the $a's$ and $b's$ are constants. \\
Then a linear transformation to normal-coordinates:
\begin{equation}
q_1 = \sum_{k=1}^n c_{ik}x_k
\end{equation}
whereby $T$ and $V$ are diagonal. The $\lambda$'s are the $n$
roots, distinct or multiple, of the determinant:
\begin{equation}
det(a_{ik}-b_{ik}) =0
\end{equation}
The Hamiltonian may be then written:
\begin{equation}
H = H_1 + \, ...\,+ H_6
\end{equation}
where
\begin{equation}
H_i = \frac{1}{2} p_i^2 + \frac{1}{2} \lambda_j x_i^2
\end{equation}
So one has an aggregation of $6$ independent simple harmonic
oscillators, i.e., in the language of wave mechanics, the wave
function of the whole system is the product of the wave function
for the individual oscillators and characteristic value is the sum
of the individual eigenvalues. This method is allowed because the
system is separable in $6$ normal coordinates.

The properties of $6$ normal fundamental vibrations frequencies
related to
$\lambda$ can be obtained following the {\em Hund }'s analysis. \\
In the investigation, the molecule $NH_3$ is assumed to have a
certain geometric symmetry in its equilibrium configuration. In
fact in considering the vibration spectrum of a tetratomic
molecule of the general type $XY_3$ (i.e. $NH_3$), the assumption
is that in the normal state of the molecule the $X(N)$-atom at

equilibrium position is equidistant, i.e., at the center of
gravity, from each of $Y(H)$-atoms which themselves lie at the
corners of an equilateral triangle, not in the same plane in which
$X$-atom is. So a regular pyramid is the normal configuration of
$NH_3$. The approximation is that the force fields between the
$X$-atoms is strong and those connecting the $X$ and the $Y$ atoms
is weak. In this case the potential function energy is assumed to
have the same symmetry as the geometric configuration of the
molecule. Then will be two frequencies $\nu_1$ and $\nu_2$
corresponding to the mutual vibrations of the $Y_3 (H_3)$ group
alone which have just the properties of triatomic molecule
\cite{De}. In $\nu_1$ the $Y$ atoms remain at the corners of an
equilateral triangle throughout the motion. This oscillation is
along the symmetry axis so it is called a $||$ vibration. While
$\nu_2$ is a double frequency due to an isotropic vibration of
$N$-atom in a plane perpendicular to the symmetry axis, it is a
$\perp$ vibration. The remaining normal vibrations of the system
may be determined by considering the motion of the $Y_3$ group,
taken as a rigid triangle, relative to the $X$ atom. The vibration
will consist of two sorts, a vibration $\nu_3$ in which the
triangle and the point $X$ oscillate with respect to each other,
the triangle plane remaining always parallel to itself. Then
$\nu_3$ is a single and a $||$ vibration. The last frequencies
$\nu_4$ is represented by a typing motion of the triangle relative
to the $X$-point. It is a double $\perp$ vibration frequency. So
there are four independent active frequencies, two $||$ and two
$\perp$. Since the latter are double, there are six degrees of
internal freedom corresponding to the formula of internal degrees
of freedom for four atoms we have seen $n=3s -6$ where $s$ is the
number of atomic nuclei which one
assumes to have a possible equilibrium position.\\
So this qualitative discussion done by Dennison, Hund allowed them
to predict the essential features of IR spectrum of the $XY_3$
molecule. There will be four fundamental absorption bands. The
intensity will be different depending upon the force fields, i.e.,
the configuration of the molecule. The fine structure of the band
$\nu_1$ is similar to the fine structure of the band $\nu_3$ since
they both correspond to a vibration along the symmetry-axis. The
pair of bands $\nu_2$ and $\nu_4$ will have a similar fine
structure because $\perp$ to the symmetry axis and will be unlike
to the pair $\nu_1$ and $\nu_3$. Questions with regard to fine
structure arise when one discusses experimental spectra by
spectrometer analysis.

\subsection{Ammonia molecule analysis with Theory of Groups}
The ammonia infrared spectrum is an example of the application of
group theory \cite{group} to physics. Molecules absorb and emit
electromagnetic radiation in wide areas of the spectrum. If
electrons change state, the radiation may be in the visible
region. Molecular ultraviolet spectra are rather rare, since
molecules fall apart at these high energies. Changes in
vibrational states are associated with infrared wavelengths, and
changes in rotational states with the far infrared. There are even
finer energy differences that cause spectra even in the
radio-frequency region. All of these generally consist of a great
number of lines, sometimes not resolved individually, forming
bands and such.

Infrared spectra are a valuable tool for determining the structure
of molecules. 
An infrared band is simpler than the band spectra in the visible,
but still rather complex, consisting of several series of lines
corresponding to transitions between different rotational states.
Two methods are generally used, absorption spectra that study the
transitions from the ground state to excited states, and Raman
spectra that studies the changes in wavelength in scattered
radiation. Raman spectroscopy can be done in the visible region
with its more convenient experimental conditions, and with the
powerful beams of lasers.

Quantum mechanics is necessary for the understanding of molecular
spectra, which it perfectly explains. Then there is a relation of
group theory to quantum mechanics. Symmetry is a powerful tool in
the quantum mechanics of molecules, and the ammonia molecule
furnishes a good example. One can  consider what infrared and
Raman spectra are to be expected if the molecule is a symmetrical
pyramid, which is indeed the case using the character analysis.
The symmetrical pyramid has the symmetry group $C_{3v}$, whit its
character table\footnote{This table defines the abstract group Ci,
which has many representations, or concrete realizations. Let the
symbol $\sigma$ stands for the transformation x = -x. C is either
of the rotations. In three dimensions, this would be a reflection
in the yz-plane. We can use $\sigma$ as an operator: $\sigma f(x)
= f(-x)$, $E$ is the identity operator, such that $Ef(x) = f(x)$
for any f(x).One calls the elements $E$ and $A_i$, that all obey
the same multiplication table.}:
\\
\\
\\
\begin{tabular}{ccccc}
\hline $C_{3v}$ & $E$ & $2 C$ & $3\sigma$ & $basis functions$
\\ \hline
 $A_{1}$ & 1 & 1& 1 & $T_z, x^2 + y^2, z^2$
\\
 $A_{2}$ & 1 & -1& 1 & $R_z$
\\
 $E$ & 2& -1& 0 & $(T_x, T_y)(R_x, R_y)(x^2-y^2, xy)(xz, yz)$
\\
\hline
\end{tabular} $\!\!\!\!\!$
\\
\\
\\
\\
\\
T and R are the representations to which components of the
translation and rotation displacements belong; these are vectors
and axial vectors, respectively. T also shows the representations
of the dipole moment operator which produces the infrared
spectrum. Then there are the quadratic functions which transform
like the molecular polarizability, the operator which produces the
Raman spectrum.

We assign three displacement coordinates to each atom, $12$ in all
for the four atoms. The first thing to do is to find the
characters of this representation. The character for E is 12,
since the identity transforms each coordinate into itself. The
rotations about the axis leave only the displacements on the
nitrogen in the same place, and the character is the same as that
of the three T components, or 1 - 1 = 0. Reflections in a vertical
plane leave the nitrogen and one hydrogen unmoved, and the
character is easily seen to be 2 - 1 = 1 for each atom. Therefore,
the characters of the reducible representation of the
displacements is 12, 0, 2. This must include the representations
of the translation and rotation of the molecule as a whole, $A_1 +
A_2 + 2E$. Therefore, we subtract the characters $6, 0, 0$ to find
the character of the vibrations, $6, 0, 2$. By character analysis,
we find that this gives $2A_1 + 2E$. Ammonia, therefore, should
exhibit four fundamentals, all active in both infrared and Raman
spectra. This is exactly what is observed. The Raman spectra of
the E fundamentals ought to be faint, and they were not observed
(or were not until lasers came in). If the ammonia molecule were
planar, two more fundamentals would be expected, and they are not
observed.
Herzfeld gives the four modes as follows. There is a
very strong band at $1627.5 cm^{-1}$ (infrared spectroscopists use
the reciprocal of the wavelength, since it is proportional to the
frequency and the quantum energy), about $6.1 \mu$, and is a
so-called perpendicular band, which would be expected from the x
and y components of the dipole moment. This is one of the
doubly-degenerate E fundamentals, a symmetric bending of two of
the hydrogens to or away from each other. The asymmetric bending
is of higher frequency, $3414 cm^{-1}$, and difficult to observe.
These are the two E modes. There is a strong parallel band at
$931.58 cm^{-1}$ and $968.08 cm^{-1}$, about $10.6 \mu$
corresponding to an $A_1$ representation. This band is double, and
the reason is curious. The ammonia molecule can turn itself
inside-out; that is, the nitrogen can pass through the plane of
the hydrogens. This isn't easy, but the nitrogen can tunnel
through, and the doubling is the result. The states divide into
those symmetrical with respect to this inversion, and those that
are antisymmetrical (change sign). The selection rules on the
rotational transitions make the band separations the sum of the
inversion splitting in the two cases. In the Raman spectrum, the
separation is the difference of the splitting. The Raman bands are
observed at $934.0 cm^{-1}$ and $964.3 cm^{-1}$. Finally, there is
a strong band at $3335.9 cm^{-1}$and $3337.5 cm^{-1}$, and a Raman
shift at $3334.2 cm^{-1}$ (about $3.0 \mu$) corresponding to the
other $A_1$ fundamental. In this mode, the bond lengths lengthen
and shorten symmetrically. The two A modes can be called bending
and stretching, respectively.

$ND_3$, with the heavier deuterium substituted for the protons,
gives somewhat different (lower) frequencies, and the shifts can
be used to nail down the identification of the vibrational
frequencies, confirming the conclusion that ammonia is a
symmetrical pyramid. The inversion doubling is a very interesting
phenomenon. It turns out to be possible to separate molecules in
even and odd inversion states, and this led to the ammonia maser,
the first of its kind. Although one can form a good picture of
ammonia as if it were a macroscopic object, try to picture it with
the nitrogen partly on both sides of the hydrogens!
\\
Using the group theory {F. Hund} (1925) \cite{[FN32]} studies the
equilibrium of the molecule of ammonia, and he shows that, if the
electronic configuration around the nitrogen, originally central,
is capable of a polarization induced by the hydrogen nuclei, the
molecule in the normal state have just the axial symmetrical form.
So he assumed that the molecule $NH_3$ has a regular pyramid
equilibrium configuration. The nitrogen atom at equilibrium
position is equidistant, i.e. at the center of gravity, from each
of hydrogen atoms, which lie at the corners of an equilateral
triangle, not in the same plane of $N-$atom. In considering the
vibrations of such a molecule he erroneously states that there
exist only three active characteristic frequencies, whereas,
unless the particles all lie in the same plane, there must in
general exist four, as shown by Dennison \cite{[FN33]}. Hund gives
a table of harmonic and combinations bands of $NH_3$ with three
fundamental frequencies $\nu_1= 970 cm^{-1}, \nu_{2}= 1700
cm^{-1}, \nu_{3}= 4500 cm^{-1} $, that may be changed by allowing
the band at $\lambda = 97\mu$ i.e. $\nu \sim 3300 cm^{-1}$ to
become the fourth fundamental band.

\subsection{Approximate analysis of $NH_3$ with WKB method.}
Dennison and Uhlenbeck \cite{du} compute the level separation of
$NH_3$, using the Wentzel-Kramers-Brillouin (WKB) method of
approximation for a one-dimensional system of a particle moving in
a potential field, consisting of two equal minima. Then they make
an application of the results to the ammonia molecule to determine
its form. The WKB method yields an approximate solution of the
wave equation whose form depends upon whether the region
considered lies within or without the region of classical motion,
that is, the region where the kinetic energy is positive. In the
first case the solution is oscillatory, in the second or
non-classical region the solution consists of a linear combination
of an increasing and a decreasing exponential. At each boundary or
critical point are valid the so-called Kramers connection formulae
\cite{kr}. These formulae furnish a method by which one may
approximate to any solution of the wave equation.\\
They show that the infrared spectrum of the ammonia molecule
exhibits features which may be directly related to the one
dimensional problem of two equal minima. The parallel type
vibration bands for example are observed to be composed of two
nearly superimposed bands, depending upon the fact that there are
two equivalent positions of equilibrium for the nitrogen nucleus.
Symmetrical molecules of the $NH_3$ type which are not coplanar
exhibit that all vibrational levels ar double, depending upon the
fact that there are two exactly equivalent positions of
equilibrium for $N$ atom, one above the plane of the $H$ atoms,
and the other at an equal distance below. A quantum mechanical
treatment reveals that it causes the vibrational level become
double. The doublet separation is small compared with the spacing
of vibrational levels (inversion problem related to rotational
spectrum).

The physical origin and theoretical description of this doubling
is presented, followed by a description of the experimental
measurement. The inversion doubling of about 35 cm-1 represents an
excellent coupling of a simple infrared measurement with a quantum
mechanical description involving many aspects of the wave nature
of vibrations. The normal modes of Ammonia are $\nu_2= 950
cm^{-1}$,(symmetric bend), $\nu_{4a}= 1627 cm^{-1}$ (asymmetric
bend), $\nu_{4b}= 1627 cm^{-1}$ (asymmetric bend), $\nu_{1}= 3336
cm^{-1}$ (symmetric stretch), $\nu_{3a}= 3414 cm^{-1}$ (asymmetric
stretch), $\nu_{3b}= 3414 cm^{-1}$ (asymmetric stretch);
$\nu_{3a}, \nu_{3b}$ are degenerate modes, as are $\nu_{4a},
\nu_{4b}$. All six normal modes are IR active.

\subsection{Exact analysis of $NH_3$.}
Rosen and Morse \cite{mors} give an analysis of the vibration of
the nitrogen in the ammonia molecule using an exact solution of
the wave equation for a form of one-dimensional potential energy.
The potential energy for this molecule has two minima at distance
$2x_m $ apart, separated by a "hill" of height $H$.\\
They describe another solution, for a form of potential field
different from that of Dennison \cite{De} and they give an example
of its application to the vibrational states of $NH_3$. Due to the
symmetry of the molecule there are two equivalent positions of
equilibrium for the nitrogen, at equal distances above and below
the plane of the three hydrogens. This equivalence of the two
minima makes every vibrational level a doublet, a result which is
found experimentally. To analyze the vibrational behavior one
separates off the coordinates of the center of the gravity of the
molecule and the Euler angles fixing its orientation in space, and
deal only with the coordinates fixing the relative positions
of the atoms.\\
One of these coordinates is $x$, the distance of the nitrogen
($N$) atom from the plane of the hydrogen. The other five
coordinates $z_1, z_2, z_3, z_4, z_5$, can be chosen that the
positions of the two equilibrium configurations are at $z_1=z_2=
z_3=z_4=z_5=0, x=\pm x_m$. The potential function $V(x,z_1, z_2,
z_3, z_4, z_5)$ therefore has its two minima at these two points.
They justify the use of $x$ as a "normal" coordinate (i.e.
splitting from the general six-dimensional problem to a
one-dimensional problem in $x$ alone) by the following method.
From considerations of symmetry all the wave functions are
symmetric or antisymmetric about the nodal hypersurface $x=0$.
They give a two minima potential field $V(x)$ which is amenable of
exact solution. For each level of the one minimum problem there is
a pair of levels for the double minimum case. The separation
between the levels in a pair is small compared to the energy
difference between different pairs as long as the levels are below
the top of the intermediate hill.

Salant and Rosenthal in 1932 \cite{ros} derive expressions for the
effects of isotopy on the normal frequencies, following Dennison's
\cite{De} general, noncentral force treatment of the normal modes
of vibration of symmetrical triatomic and tetratomic molecules.
\\
Sanderson and Silverman in 1933 \cite{sils}, following the
procedure of Dennison \cite{De}, calculate the positions of the
fundamental vibrations of molecule $ND_3$.
\\
Rosenthal \cite{ros1} summarizes briefly the general procedure for
obtaining the normal vibration frequencies of a molecule of any
type of symmetry, without the use of group theory. He writes the
expression for the kinetic energy $T$ in terms of the
displacements of the various atoms from their equilibrium
positions. The potential energy, $V$, is written in terms of the
mutual displacements of the atoms as the most general quadratic
form consistent with geometrical symmetry. As the next step,
linear combinations of the original displacements are introduced
and both $T$ and $V$ are transformed to them. The normal vibration
frequencies, $\omega$, or rather $\lambda= 4\pi^2\omega^2$ are
then obtained as the roots of $|\lambda T -V| =0$. For $n$ degrees
of internal freedom, the expansion of this $nth$ order determinant
will give rise to an equation in $\lambda$ of the $nth$ degree. He
gives a discussion of the vibration frequencies and isotopic
shifts of tetratomic molecules, with a discussion of various
intramolecular forces and the physical meaning of the results, for
pyramidal and coplanar molecules.
\\
Manning \cite{mac} chooses an expression for the potential energy
of $NH_3 (ND_3)$ which has the correct general characteristics of
geometry symmetry of $NH_3$ and which permits an exact solution of
the Schrodinger equation. Making substitutions they obtain the
indicial equation from Schrodinger equation and make quantitative
calculations of the behavior of the energy levels, those below the
top of the center of the hill of $V$ are double according data
(Wright, Randall \cite{Ra}).
\section{On the oscillations bands of ammonia by Majorana}
Majorana studied the $NH_3$ spectra \cite{vol} and obtained
results in agree with the experimental results, i.e., two simple
vibrations  and two double vibrations. He considered the symmetry
of the $NH_3$. \noindent The three atoms $H$ occupy the vertices
of equilateral triangle; the atom $N$ is on the axes out of the
plane. The independent displacements which contribute to elastic
forces are six and they obtain from the twelve displacements of
the four atoms with the condition that the resultant of applied
vectors $\delta P_i$ at the rest points $P'_i$ is zero.

He defines the displacements $q_1 = 1$, $q_2 = q_3 = \ldots = q_6
= 0$ as those in which the atom $H^1$ moves in direction $N H^1$
of $M_N/(M_N+M_H)$ and the atom $N$ in the opposite direction of
length $M_H/(M_N+M_H)$. Similarly one defines the displacements
$q_i = \delta_{i 2}$ e $q_i = \delta_{i 3}$. Then we define as
displacement $q_i = \delta_{i 4}$ that in which the atom $H^3$
shifts of 1/2 in the direction $H^2 H^3$ and the atom $H^2$ of 1/2
in the opposite direction; for circular permutation he puts the
displacements $q_i = \delta_{i 5}$ e $q_i = \delta_{i 6}$.

Indicating $\alpha$ the angle (in the equilibrium position)
$\widehat{N H^1 H^2}$ and with $\beta$ the angle $\widehat{H^1 N
H^2}$ the kinetic energy is:
 \bea
T &=& \frac{1}{2} \, \left[ \frac{M_H^2 M_N}{(M_N+M_H)^2} \,
\left( \dot{q}_1^2 \, + \, \dot{q}_2^2 \, + \, \dot{q}_3^2 \, + \,
2 \dot{q}_1 \dot{q}_2 \cos \beta \, + \, 2 \dot{q}_2 \dot{q}_3
\cos \beta
\right. \right. \nonumber \\
& & + \,  \left. 2 \dot{q}_3 \dot{q}_1 \cos \beta \right) \, + \,
\frac{M_N^2 M_H}{(M_N+M_H)^2} \, \left( \dot{q}_1^2 \, + \,
\dot{q}_2^2
\, + \, \dot{q}_3^2 \right)  \nonumber \\
& & + \,  \frac{M_N M_H}{M_N+M_H} \, \cos \alpha \left( \dot{q}_1
\dot{q}_5 \, + \dot{q}_1 \dot{q}_6 \, + \dot{q}_2 \dot{q}_6 \, +
\dot{q}_2 \dot{q}_4 \, + \dot{q}_3 \dot{q}_4 \, + \dot{q}_3
\dot{q}_5 \right)
\,  \nonumber \\
& & + \,  \left. \frac{1}{2} M_H \, \left( \dot{q}_4 \, +
\dot{q}_5 \, + \dot{q}_6 \, + \frac{1}{2} \dot{q}_4 \dot{q}_5 \, +
\frac{1}{2} \dot{q}_5 \dot{q}_6 \, + \frac{1}{2} \dot{q}_6
\dot{q}_4 \right) \right]. \eea then he defines the potential
energy \be V \; = \; \frac{1}{2} \, \sum_{i k} \, a_{i k} \, q_1
q_k
\end{equation}
and he performs a canonical transformation \cite{vol}. He obtains
a new expression of the kinetic energy in the new
 coordinates $Q_i$, similarly for the potential. He obtains then two simple
vibrations relative to coordinates $Q_1$ and $Q_2$ and two double
vibrations relative to coordinates $Q_3$ and $Q_4$ with the square
of angular velocity: \be \lambda \; = \; 4 \pi^2 \, \nu^2
\end{equation}

\section{Brief experimental investigation until 1932}
Now we will give a brief chronology of the experiments on $NH_3$.\\
Fox studies (1928) the IR region of the spectrum of $NH_3$ using
the Prism spectrometer \cite{fox}. Sir Robert Robertson and J.J.
Fox in 1928 used a small infra red prism spectrometer, filled of
ammonia gas. They took a source of energy constant, for
calibrating the mechanism for reading wave lengths. \\At $\Delta V
=(100 \pm 200)V $ and $T \sim 18° C$ they used Nernst filaments as
source of Radiation, since those gave the most uniform supply
having regard to the intensity at different regions of the
spectrum. There are source radiation - tubes observation -
spectrometer. As the full radiation contains light of short wave
length, it may affect chemically the gas $NH_3$ under observation.
The results is a measure of position of bands of $NH_3$ and their
intensity. He made a preparation of Ammonia generated in the
little flask A by warming a mixture of damp solid ammonia and
$50\%$ $KOH$ solution was allowed to escape at the two-way top
$x$, until samples were completely absorbed by water.
\\He
confirmed
that the view that the $NH_3$ is a tetrahedron was acceptable.\\
Rasetti \cite{ras} et al. have photographed the Raman spectra in
1929 of gaseous $CO_2, NO_2, NH_3$, $CH_4, C_2H_4$ using the line
$\lambda =2536$ of mercury as the exciting radiation. They have
observed vibrational transitions in all the gases, and rotational
transitions in the $NH_3$ and $CH_4$.\\
Berker in 1929 \cite{ber} analyzes the $NH_3$ absorption band
extending at $3.0 \mu$ and $1.9 \mu$ and from $8 \mu$ to $24 \mu$
interpreting the double character of the $10 \mu$ band to be a
consequence of the close proximity of the two equilibrium
positions for the $N$ atom, one of either side of the plane formed
by the $H$ atoms.\\
Dennison and Hardy in 1932 \cite{dh} make an experimental search
for the doubling of the $3.0 \mu$ band using an IR spectrometer of
high resolving power. The experimental results furnishes a strong
argument for the theory of the doubling of the ammonia bands. They
discuss the form of ammonia molecule with the theory . And then
they prove that those states of ammonia existing in nature have
vibration-rotation-nuclear spin wave functions which are
antisymmetrical for an interchange of two of the hydrogen atoms.
 \\


\section{Conclusions}
In this paper we have depicted the genesis and the first
developments of the study of spectra of $NH_3$ analyzed for the
first time by Coblentz. Far from being complete, our account has
focused on the results achieved from 1905 to 1932, as given
evidence by many articles published in widespread journals. We
have also pointed out the practically unknown contribution to
spectral analysis Majorana, who was introduced to the subject by
studies and experiments in Rome. The result reached by Majorana as
early as in the beginning of 1930 is to find the right number of
fundamental frequencies of spectrum of $NH_3$. Wide room has been
made to different approaches to study the spectrum qualitatively
and quantitatively and experimentally too. A theoretical analysis
of ammonia spectra has been reported in Sect. 2, with a brief
account of Theory of Vibrations and Theory of Groups. In the same
sections we have showed an approximate analysis of $NH_3$ with WKB
method and an exact analysis with a particular form of potential.
Particular attention has been given to the approach of Majorana
for the analysis of IR spectra of $NH_3$ in Sect. 3. Early
experiments of ammonia, essentially dealt with atomic
spectroscopy, have been discussed above in Sect.4. From what
discussed here, it is then evident the interest to study the
ammonia spectra by Majorana and its contribute to find the exact
solution.

 \vspace{2cm} \noindent {\Large \bf
Acknowledgments}

\vspace{1truecm}

\noindent The author is indebted with S.Esposito for fruitful
discussions.

\end{document}